\documentclass[a4paper]{jpconf}
\usepackage{graphicx}
\begin{document}
\title{Inverting the mass hierarchy of jet quenching effects with prompt $b$-jet substructure}

\author{Ivan Vitev}

\address{Los Alamos National Laboratory, Theoretical Division, Los Alamos, NM 87545, USA}

\ead{ivitev@lanl.gov}

\begin{abstract}
In these proceedings,  we discuss the role of heavy quark mass on the formation of parton showers. Mass effects are not well understood  when parton branching occurs in nuclear matter, such as the quark-gluon plasma.  Recently, a theoretically  consistent  picture of open heavy flavor production in ultra relativistic nuclear collisions has begun to emerge based on effective theories of QCD, such as soft collinear effective theory with Glauber gluons. We show that implementation of in-medium splitting processes containing heavy quarks into next-to-leading order calculations of heavy flavor production leads to larger cross section suppression  when compared to traditional energy loss phenomenology.  To better constrain the important mass dependence of  in-medium splitting functions, we propose a new measurement in relativistic heavy ion collisions,  based on a two-prong  structure  inside a reconstructed heavy flavor jet. In the  region of jet transverse  momenta where parton mass effects are leading, we predict a unique reversal of the mass hierarchy of jet quenching effects in heavy ion  relative to proton collisions. We find that the momentum sharing distribution of prompt $b$-tagged jets is more strongly modified  in comparison to the one for light jets. The work summarized here opens new directions of research on the substructure of heavy flavor jets.  
\end{abstract}

\section{Introduction}
Understanding the production of heavy flavor and hadronic jets is crucial to test perturbative Quantum Chromodynamics (QCD) and make full use of 
the data from the world's  high energy collider experiments. These questions are even more pertinent today,  as the center-of-mass energies $\sqrt{s_{\rm NN}}$ of hadronic and heavy ion collisions continue to increase and heavy quarks, such as charm ($c$) and bottom ($b$), are copiously produced in parton showers. The fraction of jets  initiated by prompt heavy quarks is also becoming sizable, underscoring the importance of more precise theoretical control on the effects of parton mass. In heavy ion collisions,  it was suggested more than a decade ago that these mass effects should be readily 
observable as reduced energy loss of charm and bottom quarks relative to light  quarks~\cite{Dokshitzer:2001zm}.

Experimental measurements of  $D$-mesons, $B$-mesons, and $b$-quark jets relative to light hadrons and light quark jets  in ultrarelativistic nuclear collision  have not clearly established the argued ``dead cone" effect.  This puzzle has stimulated  extensive theoretical work, focusing on collisonal energy loss, heavy meson dissociation,  and heavy flavor transport.  A comprehensive review that covers theory and experimental measurements as of a few years ago is available~\cite{Andronic:2015wma}. More recently, an important effort has been made to quantify the uncertainties in the theoretical model description of open heavy flavor and the extraction of the  transport properties of the QGP~\cite{Rapp:2018qla}.

The purpose of these proceedings  is to summarize insights on heavy meson  production in heavy ion reactions  that come from high energy physics and from effective theories of QCD for heavy quarks. A few years ago, while studying the  quenching of $b$-jets, we noticed that a very large fraction of them originate from prompt hard gluons and to some extent from light  quarks~\cite{Huang:2013vaa}.  
Global analysis of $D$-meson and $B$-meson has reliably identified gluon fragmentation a major contributor to open heavy flavor production~\cite{Kneesch:2007ey,Kniehl:2008zza}. Soft collinear effective theory has been generalized to include heavy quark masses~\cite{Rothstein:2003wh,Leibovich:2003jd}, yielding SCET$_{\rm M}$. In what follows, we discuss the formal development of such ideas for heavy ion physics and their applications to phenomenology.

\section{Heavy quark splitting functions} 
\begin{figure}[h]
\begin{minipage}{20pc}
\includegraphics[width=20pc]{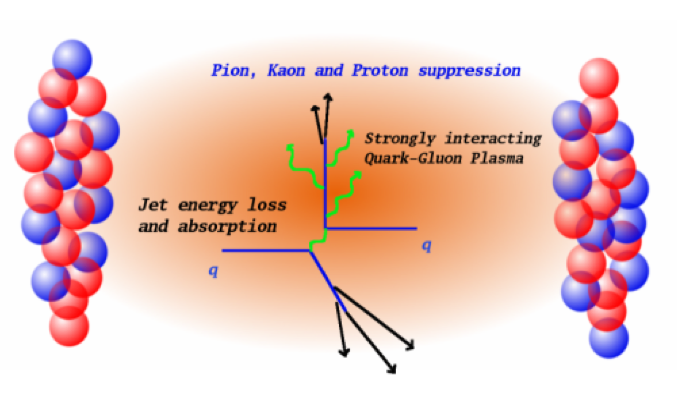}
\caption{\label{shower} Illustration of parton shower formation in the QGP produced  in
ultrarelativistic nuclear collisions. }
\end{minipage}\hspace{2pc}%
\begin{minipage}{16pc}
\includegraphics[width=16pc]{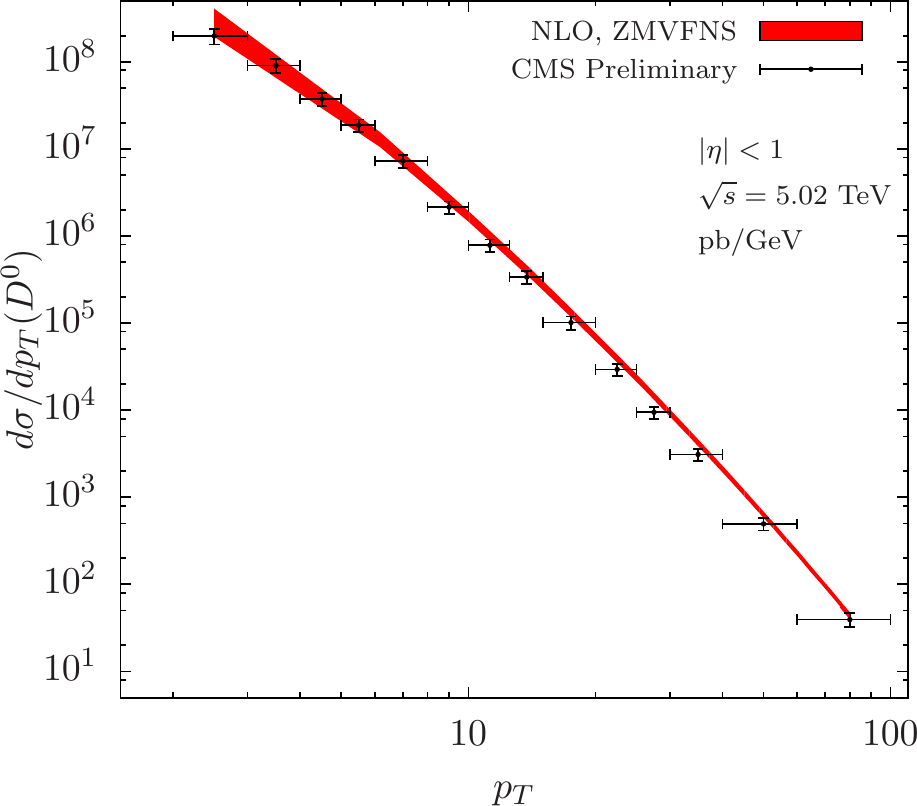}
\caption{\label{ZMFVS}  Calculation of $D^0$ production at $\sqrt{s}=7$~TeV within the ZM-VFNS scheme is compared to CMS data~\cite{CMS:2016nrh} from p+p collisions at the LHC.}
\end{minipage} 
\end{figure}

An important step toward a unified picture of particle and jet production in hadronic and nuclear collisions was the formulation of an effective theory of  QCD in the background of strongly-interacting matter. It allowed for the derivation of a full set of in-medium splitting functions for light quarks and gluons~\cite{Ovanesyan:2011kn}, the exploration of higher-order branchings~\cite{Fickinger:2013xwa} and applications ranging from an evolution approach to the quenching of light hadrons, to the suppression of light jets and their substructure modification~\cite{Kang:2014xsa,Chien:2015hda,Kang:2017frl}. See figure~\ref{shower} for a schematic of in-medium branching processes.
Recently, we performed the next logical step in this line of work by including finite mass effects in the SCET$_\mathrm{G}$ Lagrangian.  This enables the effective theory study the of interactions of heavy quarks with the QCD medium~\cite{Kang:2016ofv}. The SCET Lagrangian in the vacuum with quark masses was first derived in~\cite{Rothstein:2003wh,Leibovich:2003jd}. The corresponding theory in the vacuum is commonly referred to as SCET$_\mathrm{M}$. We labelled the new effective field theory  SCET$_{\mathrm{M,G}}$.

The recent advances in generalizing SCET$_{\rm G}$ to include finite heavy quark masses is the stepping stone for the first calculation of 
the heavy flavor splitting functions in the QGP.  We are interested in the limit $0<m < p^+$~\cite{Rothstein:2003wh,Leibovich:2003jd}, where $m$ is the
heavy quark mass and $p^+$ is the large lightcone momentum. Starting with an off-shell parton  of  momentum $[p^+ ,p^-,{\bf 0}_\perp]$  that
splits into two daughter partons $[zp^+ , {\bf k}^2_\perp/zp^+ ,{\bf k}_\perp]$ and $[(1-z)p^+ , {\bf k}^2_\perp/(1-z)p^+,- {\bf k}_\perp]$,  in the absence of a QCD medium 
 we derive the massive vacuum splitting kernels $Q\to Qg$ , $Q\to gQ$, and $g\to Q\bar{Q}$ 
\begin{eqnarray} \label{eq:Msp1}
    \left(\frac{dN^{\rm vac}}{dz d^2 \mathbf{k}_\perp}\right)_{Q\to Qg} &=&  \frac{\alpha_s}{2 \pi^2} \frac{C_F}{\mathbf{k}_{\perp}^2+z^2m^2}  
     \left( 
    \frac{1+(1-z)^2}{z}-\frac{2z(1-z)m^2}{\mathbf{k}_{\perp}^2+z^2m^2}
    \right)~,  \\
    \label{eq:Msp2}
  \left(\frac{dN^{\rm vac}}{dz d^2 \mathbf{k}_\perp}\right)_{g\to Q\bar{Q}} &=&  \frac{\alpha_s}{2 \pi^2} \frac{ T_R}{\mathbf{k}_{\perp}^2+m^2} 
    \left( 
   z^2+(1-z)^2+\frac{2z(1-z)m^2}{\mathbf{k}_{\perp}^2+m^2}
    \right)~,     \\ 
    \left(\frac{dN^{\rm vac}}{dz d^2 \mathbf{k}_\perp}\right)_{Q\to gQ} &=&       \left(\frac{dN^{\rm vac}}{dz d^2 \mathbf{k}_\perp}\right)_{Q\to Qg} (z\rightarrow 1-z) \; .
\label{eq:Msp3}
\end{eqnarray}
Here, $C_F$ and $C_A$ are Casimir operators of the fundamental and the adjoint representation of SU(3) and $T_R=1/2$. It is important to note that the longitudinal
momentum fraction $z$ and the transverse momentum ${\bf k}_\perp$ dependencies do not factorize. Furthermore, the mass term that regulates the
collinear singularity is different for the different splitting kernels in Eqs.~(\ref{eq:Msp1}) - (\ref{eq:Msp3}). Results for in-medium splitting functions 
have been obtained to first order in opacity and share similar kinematic structure  to the vacuum ones, but include the non-Abelian Landau-Pomeranchuk-Migdal
interference effects~\cite{Kang:2016ofv}.  

Heavy quark splitting functions enter parton showers and the framework of next-to-leading order (NLO) calculations. In the absence of a medium, we use the  
zero mass variable flavor number scheme (ZM-VFNS) to calculate $D-$meson and  $B-$meson production. The NLO cross sections~\cite{Jager:2002xm} 
reads
\begin{eqnarray}
\label{eq:sighadX}
\frac{d\sigma^{H}_{pp}}{dp_Td\eta}  &= & \frac{2 p_T}{s}\sum_{a,b,c}\int_{x_a^{\rm min}}^1\frac{dx_a}{x_a}f_a(x_a,\mu)\int_{x_b^{\rm min}}^1\frac{dx_b}{x_b} f_b(x_b,\mu) 
 \int^1_{z_c^{\rm min}} \frac{dz_c}{z_c^2}\frac{d\hat\sigma^c_{ab}(\hat s,\hat p_T,\hat \eta,\mu)}{dvdz}D_c^H(z_c,\mu), \qquad
\end{eqnarray}
where $\sum_{a,b,c}$ stands for a sum over all the parton flavors including light and heavy quarks and gluons. Results for $D^0$ meson production at the LHC are 
shown in figure~\ref{ZMFVS} and compared to CMS measurements~\cite{CMS:2016nrh} at the LHC.  It is important to note that $\sim 50\%$ of open heavy flavor mesons
at these energies are produced by gluon fragmentation. This leads to significantly  larger  than naively expected quenching of intermediate $p_T$ mesons in the QGP from purely
radiative processes.

\subsection{Global analysis of heavy flavor fragmentation functions and its implications for heavy ion phenomenology}
\begin{figure}[h]
\begin{minipage}{20pc}
\includegraphics[width=20pc]{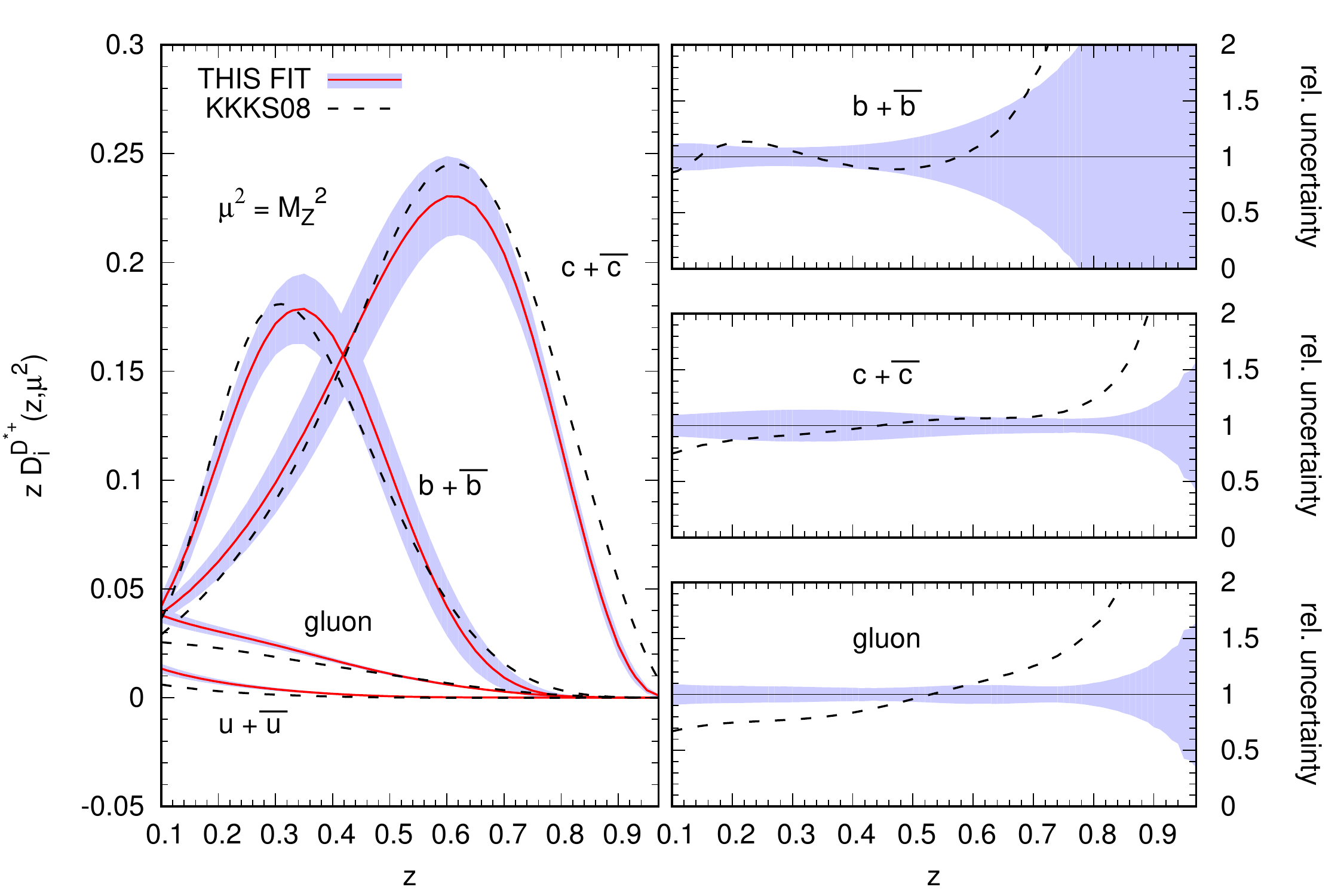}
\caption{\label{global}The fragmentation functions $z D_i^{D^{*+}}(z,\mu^2)$ 
at scale $\mu^2 = M_Z^2$ (solid red lines) along with the 
obtained uncertainty estimates (shaded bands).  
The right panels give the ratios of our uncertainty estimates (shaded bands) and
the KKKS08 fit relative to our best fit
for the $b+\bar{b}$, $c+\bar{c}$ and the gluon FF, respectively.}
\end{minipage}\hspace{2pc}%
\begin{minipage}{16pc}
\includegraphics[width=16pc]{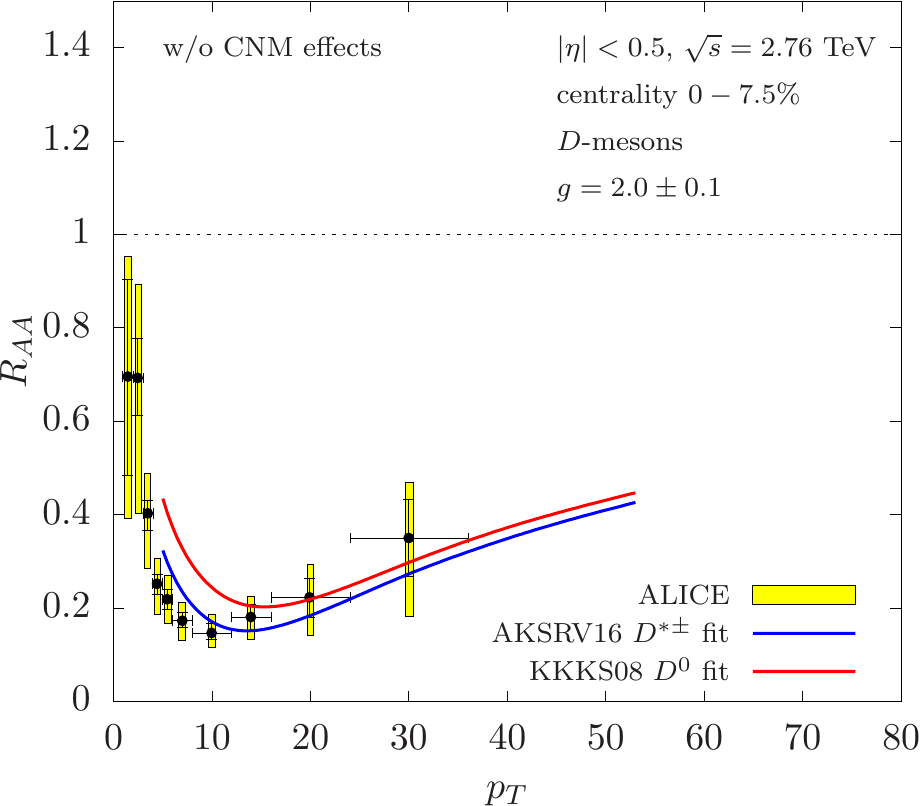}
\caption{\label{Dquench}The effect of the new AKSRV16 fit on $D$-meson quenching in heavy ion collisions at the LHC. Comparison is shown to ALICE $D$-meson suppression in 0-7.5\% Pb+Pb collisions at  $\sqrt{s_{NN}}=2.76$~TeV~\cite{Grelli:2012yv}.}
\end{minipage} 
\end{figure}

The critical importance of understanding the mechanism of open heavy flavor production has lead to novel global QCD analysis of 
charged $D^{*}$-meson fragmentation  functions (FFs) at NLO accuracy~\cite{Anderle:2017cgl}. The key advance in this work is that,  in 
addition to making use of the available data for 
single-inclusive $D^{*}$-meson production in electron-positron
annihilation and hadron-hadron collisions, for the first time in-jet fragmentation to  $D^{*}$ data~\cite{Aad:2011td} 
is included. We denote this fit as AKSRV16. We compare this fit to fragmentation functions into $D^{*+}$ mesons to the one obtained in
the KKKS08 fit~\cite{Kneesch:2007ey} in figure~\ref{global}. 
It can be seen that one of the main differences is that our fit returns a significantly larger gluon contribution, when 
compared to KKKS08 at intermediate values of $z$. One also notices, that the valence charm FFs are somewhat shifted in $z$ with respect to each other
and the height of the peak is different. The larger gluon fragmentation contribution to open charm production, however, is the most important from heavy ion physics
perspective. It implies even larger quenching of open heavy flavor.  

We investigate the implications of AKSRV16 for  the $D^*$ cross section in Pb+Pb collisions  by adding the one-loop medium correction to the vacuum NLO result. In other words, we have
\begin{equation}
\label{eq:AA}
d\sigma^H_{\mathrm{PbPb}} = d\sigma^{H, {\rm NLO}}_{pp} + d\sigma^{H, {\rm med}}_{\mathrm{PbPb}}, \quad d\sigma^{H, {\rm med}}_{\mathrm{PbPb}} = \hat \sigma^{(0)}_i \otimes D_i^{H,\mathrm{med}}.
\end{equation}
Here,  $d\sigma^{H, {\rm NLO}}_{pp}$ is the NLO cross section in the vacuum, and $d\sigma^{H, {\rm med}}_{\mathrm{PbPb}}$ is the one-loop medium correction.
We emphasize that $d\sigma^{H, {\rm med}}_{\mathrm{PbPb}}$ is negative, which leads to the quenching of the inclusive hadron cross section in heavy-ion collisions.
Phenomenological results for 0-7.5\% central Pb+Pb reactions at the LHC are shown in figure~\ref{Dquench} and compared to ALICE $D$-meson measurements at  $\sqrt{s_{NN}}=2.76$~TeV~\cite{Grelli:2012yv}.  The NLO calculation with the new AKSRV16 fragmentation functions gives noticeably larger suppression than the one with the 
KKKS08 FFs at intermediate and low $p_T$, and leaves little room for additional suppression effects. 

\section{Heavy jet substructure} 
To place stringent constraints on mass effects in parton showers, it is necessary to identify new experimental observables~\cite{Li:2017wwc}.  
The distribution of the two leading  subjets inside a reconstructed jet~\cite{Larkoski:2015lea}  is particularly suitable, since it is related to the $1\to2$  QCD splitting 
functions in the vacuum~\cite{Larkoski:2017bvj} and in QCD matter~\cite{Chien:2016led}. The jet momentum sharing variable is defined as follows
\begin{equation}
    z_g=\frac{\min(p_{T1}, p_{T2})}{p_{T1}+p_{T2}}~, \quad  z_g>z_{\rm cut} \left(\frac{\Delta R_{12}}{R} \right)^\beta \;.
    \label{zg}
\end{equation}
In Eq.~(\ref{zg}),  $p_{T1}$ and $p_{T2}$ are the transverse momenta of the subjets and soft bremsstrahlung is eliminated through the minimum  $z_g$ requirement. $\Delta R_{12}$ is the distance between two subjets and $R$ is the radius of the original jet. In  Ref.~\cite{Ilten:2017rbd} and simulations of heavy flavor
splitting functions were performed using a Monte Carlo event generator.

We are interested a in different aspect of heavy quark splitting functions -- the precise effect of mass on parton shower formation.
To prepare for the jet splitting function calculation in heavy ion collisions, we start with the vacuum case,  denote by  $j\to i \bar{i}$  
the parton branchings and define $r_g=\theta_g R$. The $\theta_g$ and $z_g$ distribution for parton $j$,  after soft-drop grooming  at the lowest non-trivial order  is  
\begin{equation}
    \left(\frac{dN^{\rm vac}}{dz_g d\theta_g}\right)_{j}  = \frac{\alpha_s}{\pi} \frac{1}{\theta_g} \sum_{i} P_{j\to i \bar{i}}^{\rm vac}(z_g)~.
\end{equation}
When the splitting probability becomes large, resummation can be performed  to modified leading-logarithmic (MLL) accuracy
and the resummed distribution for a $j$-type jet, initiated by a massless quark or a gluon,  is given by
\begin{equation} \label{eq:mll}
    \frac{dN_j^{\rm vac,MLL}}{ dz_g d\theta_g} = \sum_{i} \left(\frac{dN^{\rm vac}}{dz_g d\theta_g}\right)_{j\to i \bar{i}}
     \underbrace{\exp \left[-\int_{\theta_g}^1 d\theta \int_{z_{\rm cut}}^{1/2} dz  \sum_{i} \left(\frac{dN^{\rm vac}}{dz d\theta}\right)_{j\to i \bar{i}}  \right]}_{\rm Sudakov~Factor}~.
\end{equation}
The normalized joint probability distribution  in the domain   $z_{\rm cut} \leq z_g \leq 1/2$, $0 \leq \theta \leq 1$ can be straightforwardly obtained.

\begin{figure}[h]
\begin{minipage}{18pc}
\includegraphics[width=18pc]{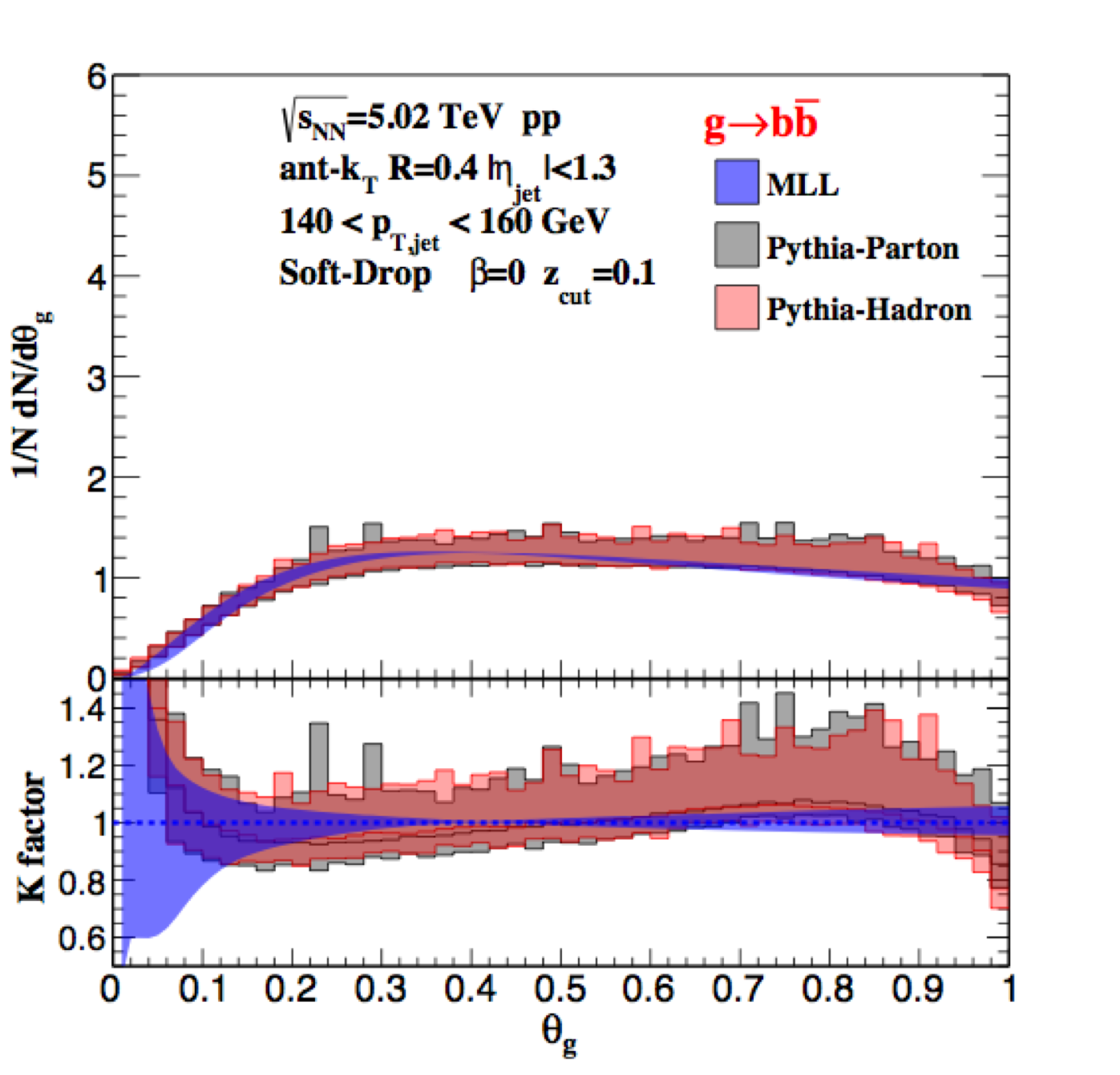}
\caption{\label{NPcorr} The MLL calculation of the angular distribution of subjets is shown versus Pythia parton level simulations, as well as simulations that include non-perturbative
hadronization effects.  
  }
\end{minipage}\hspace{2pc}%
\begin{minipage}{18pc}
\includegraphics[width=18pc]{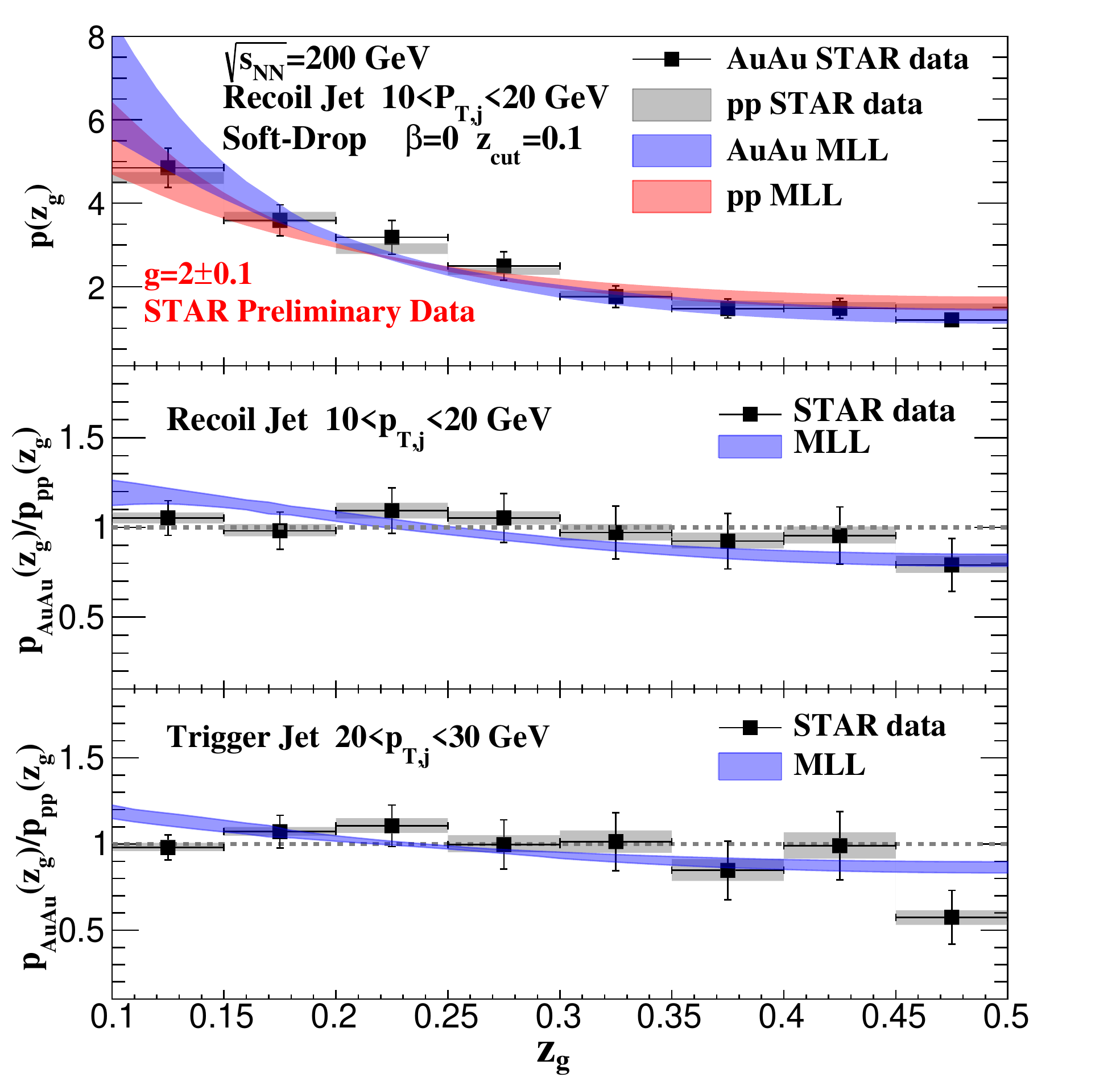}
\caption{\label{tagsplit}  Upper panel: the theoretical predictions for the  $z_g$ distribution of recoil jets to the preliminary STAR data~\cite{Kauder:2017cvz}. The middle and bottom panels show the predictions and  measurements for the modification of recoil  and trigger jets, respectively.}
\end{minipage} 
\end{figure}

 Eq.~(\ref{eq:mll}) can be extended to the 
case of prompt  heavy flavor jet splitting, such as $b\to bg$ or $c\to cg $, in a straight forward way. For gluon splitting into heavy quark pairs greater care must be taken for the probability function and it is defined as 
\begin{equation}
    p(\theta_g, z_g)\big|_{g\to Q\bar{Q}} = \frac{\left(\frac{dN^{\rm vac}}{dz_g d\theta_g}\right)_{g\to Q \bar{Q}} \Sigma_{g}(\theta_g)}{\int_{0}^{1}d\theta \int_{z_{\rm cut}}^{1/2} dz  \left(\frac{dN^{\rm vac}}{dz d\theta}\right)_{g\to Q \bar{Q}} \Sigma_{g}(\theta)  }~.
\end{equation}
Here,  $\Sigma_{g}(\theta_g)$ is the Sudakov factor for gluon evolution and it  exponentiates all the possible contributions from gluon splitting, such as $g\to g g $ and $g\to q\bar{q}$. Thus,  MLL resummation can change significantly the predictions for the $g\to Q\bar{Q}$ channel relative to the leading order (LO) results.  
Finally, the  distribution needs to be convolved with the quark and gluon jet production cross section, for which we use MADGRAPH5\_AMC@NLO. Our results for the $g\to b\bar{b}$ channel, 
are shown in figure~\ref{NPcorr}. The agreement between the theory framework and the Pythia simulations is at the 5\% level, as seen from the bottom panel  of the figure,  and non-perturbative effects are small. In contrast, for such channels  the lowest order calculation shows ${\cal O}(100\%)$ deviations.

 In the presence of a QCD medium, we replace the vacuum splitting functions by the full splitting kernels, as rigorously derived in~\cite{Ovanesyan:2011kn,Kang:2016ofv} 
\begin{equation}
    \left(\frac{dN^{\rm full}}{dz d^2 \mathbf{k}_\perp}\right)_{j} = \left(\frac{dN^{\rm vec}}{dz d^2 \mathbf{k}_\perp}\right)_{j}   + \left(\frac{dN^{\rm med}}{dz d^2 \mathbf{k}_\perp}\right)_{j}.      
\end{equation}
To reduce any model dependence, we evaluate  the QGP-induced component in the hydrodynamic background used to  describe quarkonium 
suppression~\cite{Aronson:2017ymv}. Since resummed calculations of jet splitting functions have not been performed in heavy ion collisions prior to our work~\cite{Li:2017wwc},
we start with comparison to light jets at RHIC. Results are shown in figure~\ref{tagsplit} for both the absolute momentum sharing distributions and their modification in Au+Au relative to p+p
collisions. We have compared theory to STAR preliminary measurements~\cite{Kauder:2017cvz} for both tagged and recoil jets and observe good agreement.

\begin{figure}[h]
\begin{minipage}{18pc}
\includegraphics[width=18pc]{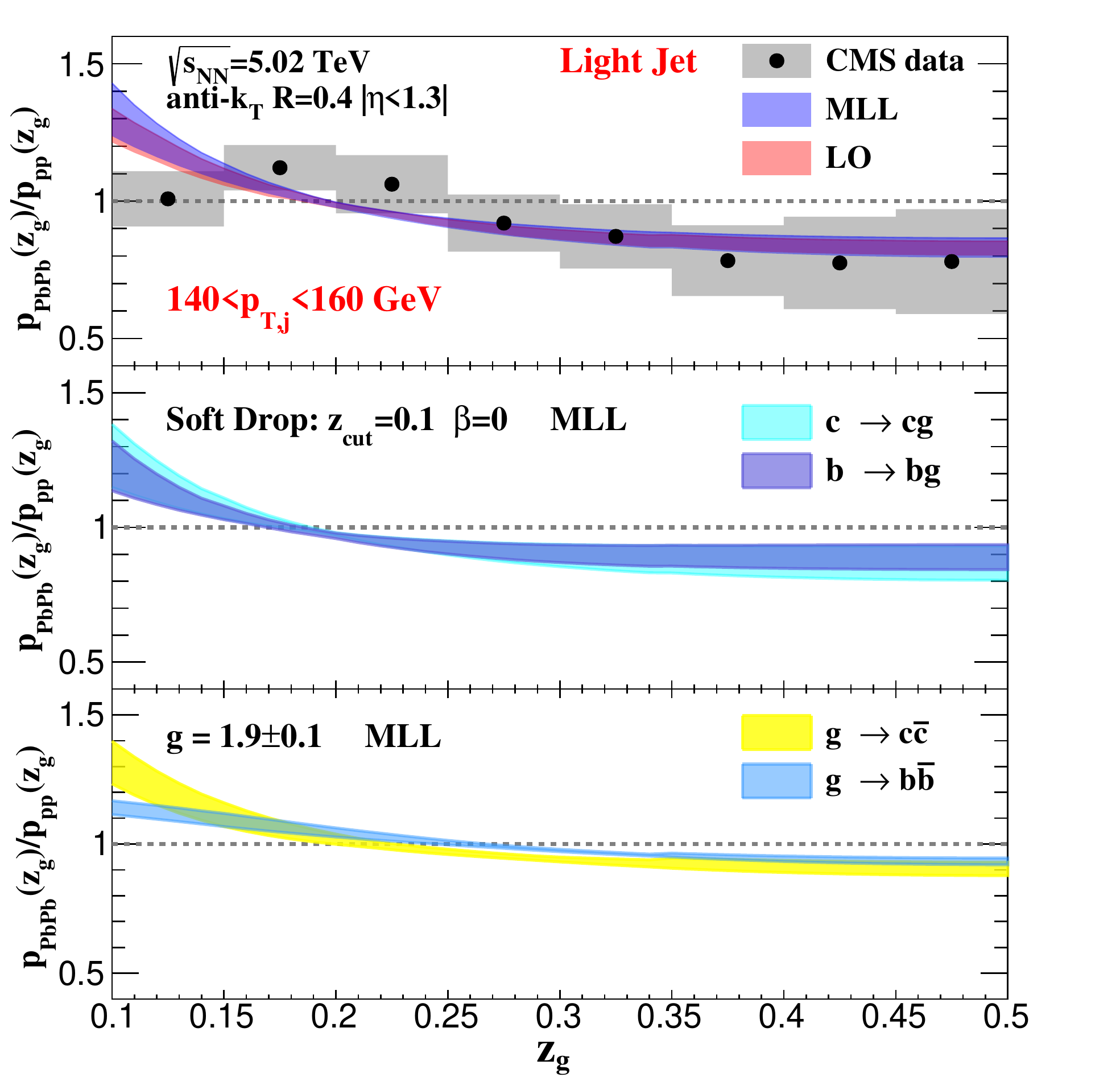}
\caption{\label{hLHC}The modification of the jet splitting functions  in 0-10\% central Pb+Pb  collisions at $\sqrt{s_{\rm NN}}=5.02$~TeV  for the $p_T$ bin  $140<p_{T,j}<160$~GeV.  The upper panels compare the LO and MLL predictions to CMS  light  jet substructure measurements~\cite{Sirunyan:2017bsd}. The middle  and lower panels 
    present the MLL modifications for heavy flavor tagged jet - the $Q\rightarrow Qg$ and $\rightarrow Q{\bar Q}$, respectively. }
\end{minipage}\hspace{2pc}%
\begin{minipage}{18pc}
\includegraphics[width=18pc]{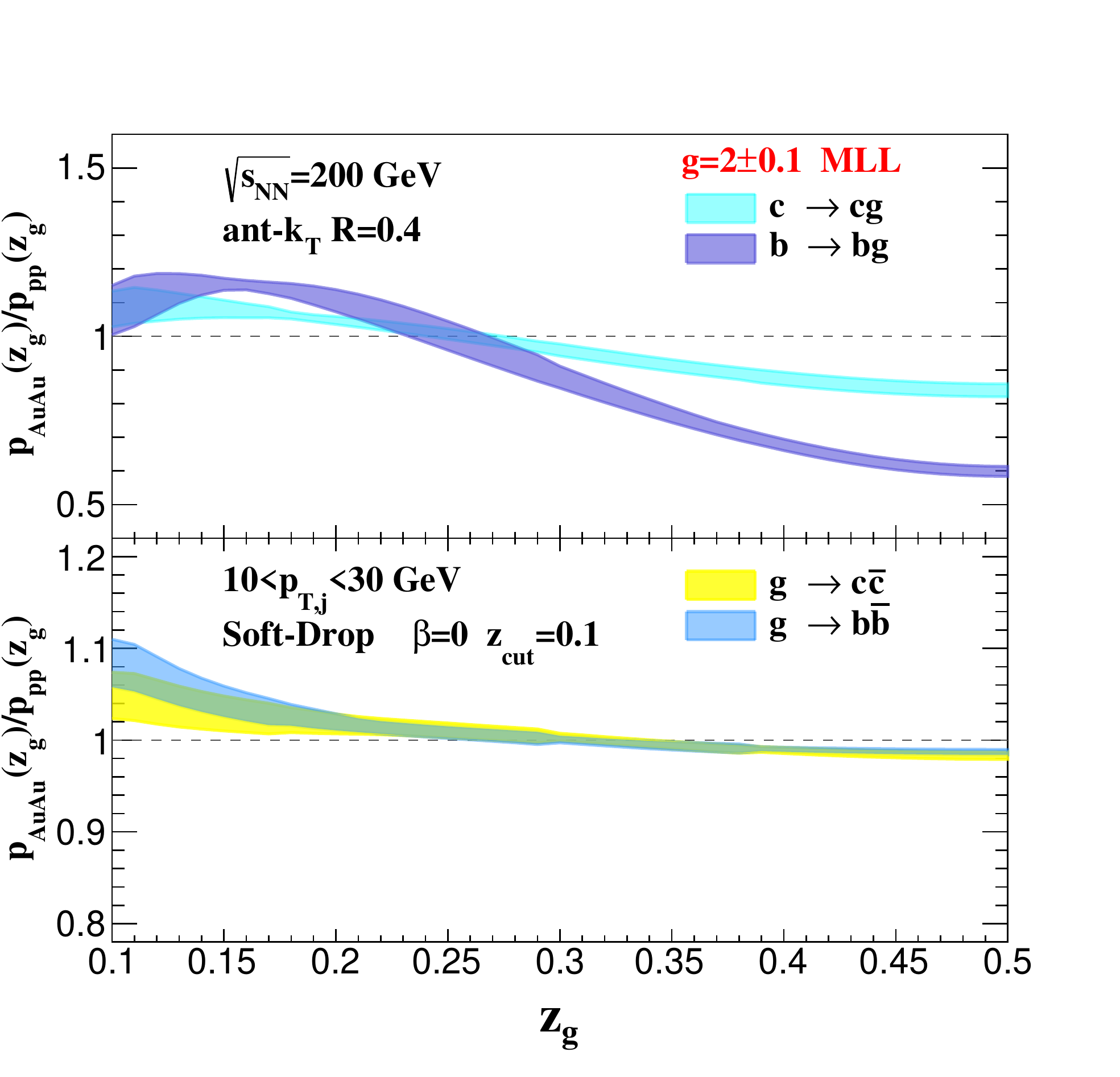}
\caption{\label{hRHIC}The modifications of the splitting functions for heavy flavor tagged jet is shown for $\sqrt{s_{\rm NN}}=200$~GeV Au+Au collisions. An important feature is the strong quenching effects for prompt
    $b$-jets contrasted by the lack of QGP-induced modification for the $g\to Q\bar{Q}$ splitting.}
\end{minipage} 
\end{figure}

We now move to the substructure of  heavy flavor jets at the LHC, with first results shown in figure~\ref{hLHC} for an illustrative kinematic range  $140<p_{T, j}<160$~GeV  in 0-10\% central Pb+Pb collisions  at $\sqrt{s_{\rm NN}}=5.02$~TeV. The top panel  confirms the good description of the CMS splitting function modification measurements~\cite{Sirunyan:2017bsd}, in this case anti-$k_T$ jets with $R=0.4$ and $|\eta|<1.3$ with the two subjets separated by $\Delta R_{12}>0.1$. 
Predictions for the modification of prompt $b$-jet and $c$-jet substructure  are given in the middle panel,  which show that jet quenching effects for $p(z_g)$ are comparable to that of light jets in this high $p_T \sim 100$~GeV region. This is not unexpected, since mass effect slowly vanishes with increasing jet energy. The bottom panel of figure~\ref{hLHC}  illustrates QGP effects for the $g \to Q \bar{Q}$ channel, where they are somewhat smaller for $g\to b\bar{b}$ in comparison to the other splitting functions.

The most important finding from our recent work~\cite{Li:2017wwc}  is that by going to lower jet transverse momenta, we can enter a regime where the leading mass dependence 
comes from the propagator in the splitting kernels. The relevant condition 
reads $\mathbf{k}^2_\perp \ll z^2_g m^2$, $\mathbf{k}^2_\perp \ll  m^2$,  $\mathbf{k}^2_\perp \ll (1-z_g)^2 m^2$  and in 
the medium induced splitting functions we typically have 2 such propagators. This leads to a unique dependence of the jet momentum sharing distributions in heavy ion collisions, 
which is different for light jets, and the different heavy flavor jet channels. 
Despite the tremendous importance of open heavy flavor production for heavy ion collisions, our theoretical expectations so far have been limited to the energy loss hierarchy  
$\Delta E^{\rm rad}_b <  \Delta E^{\rm rad}_c < \Delta E^{\rm rad}_{u,d} < \Delta E^{\rm rad}_g$ from radiatve processes in the QGP.  In contrast, we can predict analytically that the substructure 
modification will be the largest for prompt heavy flavor subjets, followed by the one for light jets. If both subjets are heavy, corresponding to $g \rightarrow Q \bar{Q}$, 
we expect no modification. This unique inversion of the mass hierarchy of jet quenching effects can be tested by experiments in the near future.   Numerical results for RHIC energies and 
momentum sharing distribution ratios for heavy flavor tagged jets in  Au+Au to p+p collisions  at $\sqrt{s_{\rm NN}}=200$ GeV are presented in figure~\ref{hRHIC} for  $10<p_{T, j}<30$~GeV  jets. 
These corroborate the analytic expectations, show that the magnitude of the effects is large, and provide new and promising ways to pin down the mass dependence of in-medium parton showers.

\section{Conclusions}
To summarize, we focused here on the effects of heavy quark mass on parton shower formation in the QGP, and the
production of open heavy flavor and heavy flavor jets in general. The recent results that we report are important steps toward
the much needed development of a unified theory that puts the description of jets and heavy flavor production in hadronic 
and heavy ion collisions on the same footing.  These advances are linked to development of  modern effective theories of
QCD in the  environment  of strongly interacting matter, such as soft collinear effective theory with Gluaber gluons.  
SCET$_{\rm G}$ has been extended to open heavy flavor, leading to the first consistent NLO calculation to  of $D-$meson
and $B-$meson suppression in heavy ion collisions. The fact that half of open heavy flavor production, especially at 
LHC energies, comes from gluon fragmentation, motivated a new global analysis of $D^*$ FFs that includes for the first time 
novel results on hadron production inside reconstructed jets. Our results show that the fraction of open heavy flavor mesons
that come from the decay of a prompt gluon is even larger than previously anticipated and their suppression through purely radiative processes in the QGP exhausts the observed 
 magnitude of jet quenching.   

We also looked for the most promising observables to accurately constrain the role of parton mass in the formation of parton showers. 
To this end, we presented the first resummed calculations of the soft-dropped  momentum sharing distributions in heavy ion collisions.   For light jets, the MLL result for this observable in Au+Au and Pb+Pb reactions agrees well with the recent experimental measurements over a wide range of center-of-mass energies. For heavy flavor tagged jets,  we demonstrated that jet splitting functions  can be used to  constrain the still not well understood dead cone effect in the QGP.  We also identified  the kinematic domain where those effects are important and predicted a unique inversion of the mass hierarchy of jet quenching effects, with the modification of the 
momentum sharing distribution for prompt $b$-jets being the largest. The work reported here sets new direction of research on heavy flavor jet substructure in ultrarelativistic  nuclear collisions and underscores the importance of future studies of 
flavor-separated jets~\cite{Chien:2018dfn}.

\ack   This work is supported on US  Department  of  Energy, 
Office of Science under Contract No. DE-AC52-06NA25396,  the DOE Early Career Program, and the LDRD program at LANL.

\section*{References}

\providecommand{\newblock}{}


\end{document}